\documentclass{emulateapj}
\usepackage{natbib}

\newcommand{\dd}{\"}

\begin{document}

\title{On the Lack of a Soft X-Ray Excess from Clusters of Galaxies}

\author{Joel N. Bregman and Edward J. Lloyd-Davies}
\affil{Department of Astronomy, University of Michigan, 830 David
  M. Dennison Building, Ann Arbor, Michigan 48109-1090}
\email{ejdavies@umich.edu}

\begin{abstract}
  A soft X-ray excess has been claimed to exist in and around a number
  of galaxy clusters and this emission has been attributed to the
  warm-hot intergalactic medium that may constitute most of the
  baryons in the local universe.  We have re-examined a study of the
  {\it XMM-Newton\/} observations on this topic by \citet{kaastra03a}
  and find that the X-ray excess (or deficit) depends upon Galactic
  latitude and appears to be most closely related to the surface
  brightness of the 1/4 keV emission, which is largely due to emission
  from the Local hot bubble and the halo of the Milky Way.  We suggest
  that the presence of the soft X-ray excess is due to incorrect
  subtraction of the soft X-ray background.  An analysis is performed
  where we choose a 1/4 keV background that is similar to the
  background near the cluster (and for similar HI column).  We find
  that the soft X-ray excess largely disappears using our background
  subtraction and conclude that these soft X-ray excesses are not
  associated with the target clusters.  We also show that the
  detections of ``redshifted'' O VII lines claimed by \citet{kaastra03a}
  are correlated with solar system charge exchange emission suggesting
  that they are not extragalactic either.
\end{abstract}


\section{Introduction}

Clusters of galaxies contain X-ray emitting hot gas (10$^{{\rm
    7}}$-10$^{{\rm 8}}$ K) that accounts for $\sim 11\%$ of the total
mass of the system and contains more baryons than the visible
galaxies \citep{allen02a, allen03a}.  There have been many studies of
this hot gas as well as searches for other cooler gaseous components
and other types of emission.  Both excess absorption and emission have
been claimed to be present in clusters, suggestive of material colder
than the ambient hot cluster material, although these have been
controversial issues.  The claims about excess absorption arose from
the {\it Einstein Observatory SSS\/} spectra (White et al.  1991),
where the soft emission was less than would be expected from Galactic
absorption of the free-free spectrum of a cluster.  These {\it SSS\/}
spectra had to be corrected for the buildup of ice in the optical
path, so there was some concern that if the correction for ice was
wrong, it might lead to the observed effect.  This result was not
confirmed with subsequent instruments, such as {\it ROSAT\/}
\citep{arabadjis00a} or {\it XMM-Newton\/} \citep{peterson03a}, so we
can safely conclude that the original study was incorrect and that
there is no substantial absorbing medium.

Substantially more controversial is the subject of an additional emission
component at soft X-ray energies (0.1-1 keV).  For many clusters, it is
claimed that the emission detected by the {\it Extreme Ultraviolet
  Explorer\/} ({\it EUVE\/}) and by four different X-ray telescopes cannot
be explained by cluster free-free emission that is absorbed by cold Milky
Way gas (e.g., \citet{lieu96a, lieu99a, lieu00a}; \citet{durret02a};
\citet{kaastra03a}, and references therein).  They argue that the emission
becomes more prominent with increasing radius from the cluster center
relative to the harder emission of the cluster, and that the temperature of
the emission is typically 0.1-0.3 keV.  They interpret this emission as
being either non thermal, due to cosmic rays in the cluster
\citep{sarazin98a} or thermal, due to gas at 1-3$\times$10$^{{\rm 6}}$ K
(e.g., \citet{kaastra03a}).  If it is thermal, its mass may be comparable
to that in the hotter ambient component, so it would have cosmological
consequences.

However, these works have been criticized for several reasons.
\citet{berghofer00a} (also, \citet{bowyer00a}, and references therein)
argued that flat-fielding corrections were not properly applied to the {\it
  EUVE\/} data, and after making this correction, no excess emission is
found, with the exception of the Coma cluster.  The {\it ROSAT\/} data were
examined by \citet{arabadjis00a} who found that the cluster spectra could
be fit with a hot free-free spectrum plus Galactic absorption and that no
additional soft component was needed (except for the Coma cluster).  The
X-ray spectra from {\it Beppo-SAX\/} was investigated by
\citet{berghofer02a}, following the study by \citet{kaastra99a} that
Abell 2199 contained a soft component.  Using a different approach to the
analysis, \cite{berghofer00a} found no evidence for an additional soft
component either in Abell 2199, or in Abell 1795.  The differences between
these works and those of Lieu and collaborators have to do with the
technical details of background subtraction and flat-fielding.

Recently, \citet{kaastra03a} used {\it XMM-Newton\/} data to search
for soft X-ray excess emission (0.2 keV) in a sample of 14 galaxies
clusters. They find evidence for excess emission in the spectra of
several of the clusters and they show that it is broadly extended
across the clusters.  They attribute this emission to the presence of
hot gas in intercluster filaments that contain the Warm-Hot
Intergalactic Medium ({\it WHIM\/}) near these clusters.  The
detection of the {\it WHIM\/} would be a major discovery and would be
best accomplished with {\it XMM\/} due to its large collecting area.
Therefore, we examine this result to understand if it is subject to
the criticism that have been raised in other observations.

\begin{figure*}
\plotone{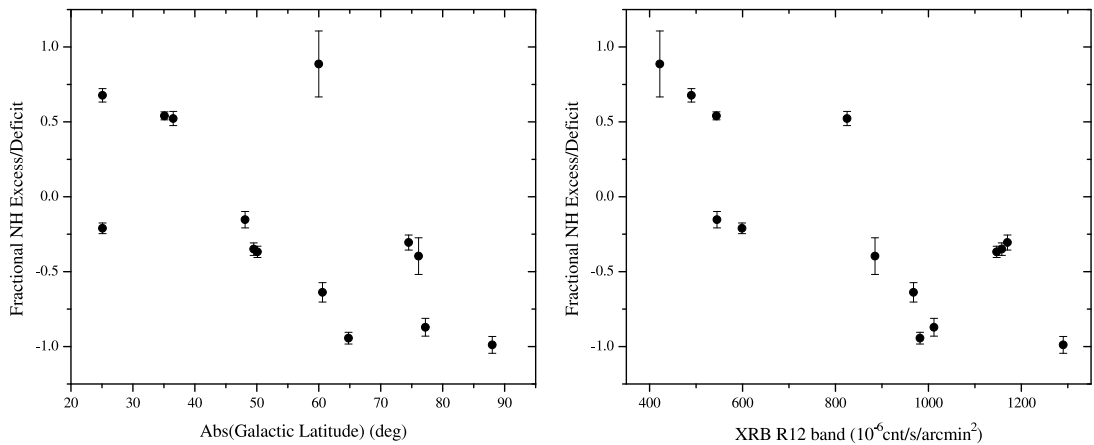}
\caption{Fractional N$_H$ excess/deficit. \emph{Left panel}:
  anti-correlation with Galactic Latitude, which is significant at the
  99\% confidence level. \emph{Right panel}: anticorrelation with 1/4
  keV X-ray intensity (R12 band from {\it ROSAT\/}) in the same part
  of the sky.\label{fig1}}
\end{figure*}

\section{Determination of the Apparent Absorption Column}

The data processing for the 14 clusters is straightforward and clearly
explained by \citet{kaastra03a} but there is a general problem that
faces observers when the size of the extended source is comparable to
the field of view.  In that case, there may not be a ``clean'' area on
the image that one can use for background subtraction.  A common
solution to this problem is to use the background from another field
(or fields), scaled to the length of exposure for the relevant
observation, and this is the approach of \citet{kaastra03a}.  This
approach works well for the high energy part of the spectrum (2-10
keV), which is due to the sum of many AGNs, and is nearly isotropic on
the sky.  However, the background at soft energies ($<$ 1 keV) is due
to the Milky Way and there is a strong latitude dependence, as well as
prominent structures around the sky due to old supernova remnants.
Kaastra recognizes the variation and adds it to the uncertainty in
extracting a flux, but as we will show, there is a systematic effect
with the background flux as a function of the soft X-ray background.

In the spectral analysis of the data, they fit a two-temperature model plus
a column density and they compare the derived column density to the
Galactic value.  They show that ten of 14 galaxy clusters have a derived
column that is below the Galactic 21 cm HI measurement in the same
directions.  The difference between the derived and Galactic columns are
often much larger than the uncertainties involved, and in some cases, the
derived column is consistent with zero.  Such low derived columns are
unphysical, so the authors argue that it is the spectrum that must be
modified.  Subsequently, they fix the Galactic 21 cm column and fit a
two-temperature model, finding a significant component at lower temperature
(0.2 keV) that is the soft excess component.

This same effect could occur if the Galactic soft X-ray background
toward the cluster was larger than the value of the mean background
field that they used.  Then, the soft component of the background
would not be fully subtracted, leaving an apparent excess to the X-ray
emission.  To test this possibility, we compare the fractional excess
or deficit in the derived absorption column with both the local value
of the Galactic soft X-ray background (R12) and with the Galactic
latitude, as the soft X-ray background is brighter toward the poles.

We calculate the fractional difference as f$_{{\rm N}}$ = (N$_{{\rm
    X}}$-N$_{{\rm 21cm}}$)/N$_{{\rm 21cm}}$, where N$_{{\rm X}}$ is
the absorption column density derived from X-ray fitting (given in
Table 2 of Kaastra et al. 2003) and N$_{{\rm 21cm}}$ is the Galactic
21 cm column.  First, we compare this quantity to the Galactic
latitude (Fig. \ref{fig1}: Left panel), which appears to show a
correlation in the sense that the higher latitude sources have
preferentially low values of N$_{{\rm X}}$ relative to N$_{{\rm
    21cm}}$.  The correlation coefficient for this relationship is
-0.65, which is significant at the 99\% level for 14 data points.

A similar and possibly better relationship exists between f$_{{\rm
    N}}$ and the soft X-ray background in the R12 band, based on the
work of \citet{snowden97a} and obtained through the tool from the High
Energy Archive (HEASARC).  Since some of the emission in this band may
be due to the cluster under consideration, we obtained values for R12
from regions $\pm5^{\circ}$ away from the cluster at constant Galactic
latitude (except for the Virgo cluster, where we took backgrounds
$10^{\circ}$ away and Coma, which we took $5^{\circ}$ away across the
Galactic North pole) which corresponds to 19(z/0.05) Mpc from the
cluster (for H$_{{\rm o}}$ = 70 km sec$^{{\rm -}{1}}$ Mpc$^{{\rm
    -}{1}}$), about an order of magnitude larger than the virial
radius of a typical cluster (2-3 Mpc).  The resulting correlation
between f$_{{\rm N}}$ and R12 (Fig. \ref{fig1}: Right panel) appears
to have less scatter than the correlation with latitude and the
correlation coefficients is -0.73, corresponding to a significance for
the relationship at the 99.7\% confidence level (when we use the R12
flux with an offset of $2^{\circ}$ toward the Galactic equator, the
significance is slightly higher, 99.9\%) .  These correlations suggest
that the presence of a soft component, preferentially for the
high-latitude sources, may be due, at least in part, to the
subtraction of the background in the soft energy band.  If correct,
this might be evident from the positions of the clusters on the soft
X-ray background.

\begin{figure*}
\plotone{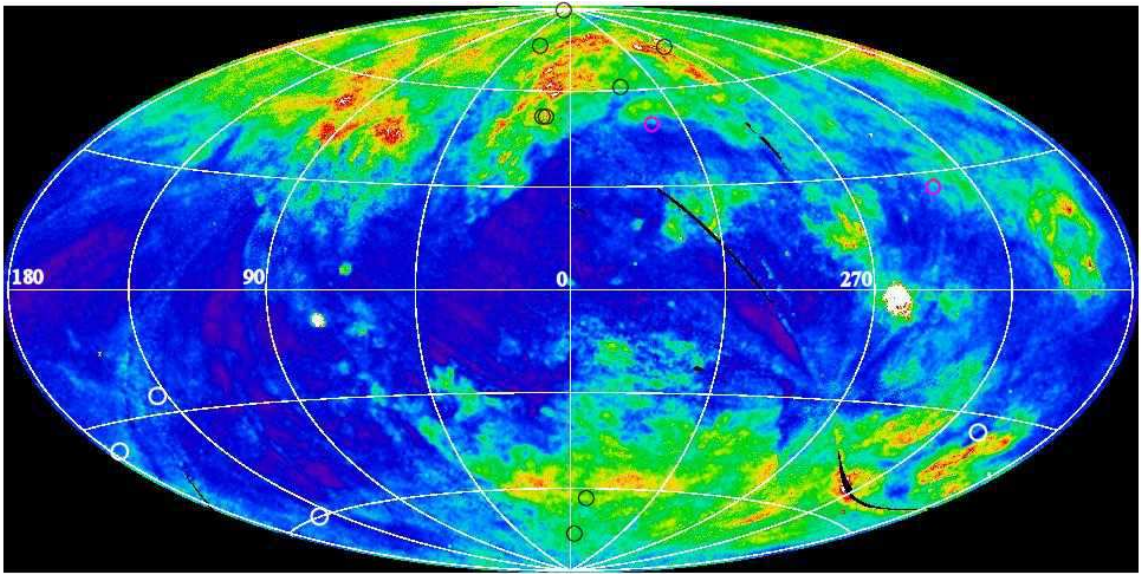}
\caption{The all-sky image of the 1/4 keV X-ray intensity (R12 band
  from {\it ROSAT\/}), along with the locations of the galaxy clusters
  analyzed by \citet{kaastra03a}.  The black circles have an apparent
  N(HI) deficit (X-ray excess objects) and are located in regions of
  higher than average X-ray brightness (note that two clusters lie at
  about {\it l\/} = $10^{\circ}$ , {\it b\/} = $50^{\circ}$ ).  The
  white circles have an apparent N(HI) excess and generally lie in
  regions where the background intensity is lower than average.  The
  purple/grey circles are the two clusters with neither an excess or
  deficit.\label{fig2}}
\end{figure*}

The all-sky 1/4 keV map (R12) has a great deal of structure due to
well-known features, such as the North Polar Spur, and there is a
brightening toward the poles due to the presence of a Galactic halo
(0.1-0.2 keV) along with the Local Bubble of hot gas (0.1 keV), which
is probably elongated toward high Galactic latitudes (Fig.
\ref{fig2}).  Upon this figure, we show clusters that have too little
absorption (f$_{{\rm N}}$ $<$ -0.3; the soft excess objects), excess
absorption (f$_{{\rm N}}$ $>$ 0.3), and those consistent with Galactic
absorption.  Most of the objects with f$_{{\rm N}}$ $<$ -0.3 lie in
regions of enhanced emission in the map and these regions are often
part of larger structures.  For example, MKW3s, Abell 2052, the Virgo
cluster, Abell 1795, and Abell 1835 lie on or very close to the North
Polar Spur, an old superbubble.  The cluster Abell S1101 (also know as
Sersic 159-03) lies on the edge of a large bright region toward the
Southern Galactic Pole and Coma covers the North Galactic Pole,
another large bright region (more on Coma below).  In contrast, all
four objects with f$_{{\rm N}}$ $>$ 0.3 (excess absorption) lie in
regions of low diffuse X-ray emission (Abell 496 is just a few degrees
away from a bright ridge).  One of these four objects, the NGC 533
cluster, is at high Galactic latitude ($-60^{\circ}$) and with same 21
cm column as toward the high latitude sources MKW3s and Abell 2052
({\it b\/} = $50^{\circ}$; both are excess emission objects), the
primary difference being the values of the Galactic soft X-ray
background.

\begin{table*}
\begin{center}
  \caption{Properties of the four re-analyzed cluster fields in
    order of increasing galactic N$_{21cm}$. The R12 values are the mean of
    four points, 2 degrees away from the cluster center in different
    directions. \label{tab1}}
\begin{tabular}{lccccccc}
\tableline\tableline
Cluster & OBSID & l & b & z & N$_{21cm}$ & R12 & Exp. \\
& & (deg.)& (deg.)& & ($10^{20}$ cm$^{-2}$)& & (ks)\\
\tableline
Abell 1795& 0097820101 & 33.7876& 77.1553& 0.06248& 1.17& 1076.5& 34.66\\
Abell S1101& 0123900101 & 348.3422 & -64.8125& 0.05800& 1.83 & 1031.8 & 27.90\\
Abell 1835& 0147330201 & 340.3759& 60.5878& 0.25320& 2.28& 1025.6& 36.93\\
MKW 3s& 0109930101 & 11.3938 & 49.4583& 0.04500& 3.04 & 1118.8 & 31.05\\
\tableline
\end{tabular}
\end{center}
\end{table*}

\section{Consequences of Matched Background Subtraction}

If the inference that the presence of the soft excess is related to
the removal of the soft background is correct, then the soft excess
should be reduced or vanish when a more appropriate background is
used. The standard method used for background subtraction of clusters
with {\it XMM-Newton\/} is to use a ``blank sky'' background created
by taking a number of observations that do not contain extended
sources or bright point sources, and stacking them together to obtain
a mean background. Simple prepackaged backgrounds for the EPIC-pn and
mos instruments have been produced by \citet{lumb02a} and
\citet{read03a} have produced an extensive set of backgrounds for each
combination of instrument mode and filter. However these are at best
mean backgrounds, averaged across the sky, and are in reality the mean
of a small number of biased sky positions whose relation to the real
all-sky mean is undetermined. Since the low energy end of the X-ray
background is the most spatially variable, due to the galactic
background and absorption, the errors introduced by the use of these
mean backgrounds will be largest at low energies.

\begin{figure*}
\plotone{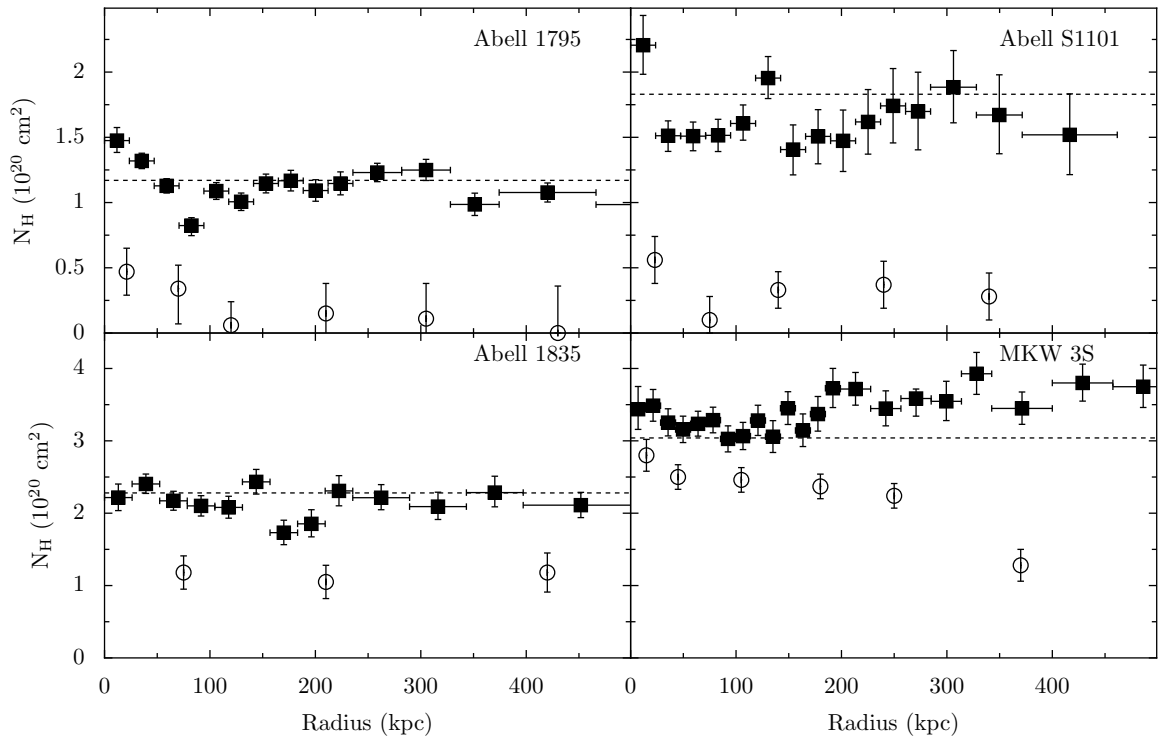}
\caption{Comparison of the fitted N$_H$ obtained by \citet{kaastra03a}
  (using an mean ``blank sky'' background) for four clusters (Abell
  S1101, Abell 1835, Abell 1795 and MKW 3s) (open circles) with those
  obtained using backgrounds matched to properties of the individual
  fields (solid squares).\label{fig4}}
\end{figure*}

A background drawn from ``blank'' fields with similar properties to
those of the target cluster fields, rather than random ones, should
match the cluster backgrounds much more closely. In order to test this
we constructed ``blank sky'' backgrounds matched to properties of the
individual cluster fields. The technique we used to accomplish this
involves identifying three parameters that are likely to affect the
observed background: N$_{\rm 21cm}$, R12 and particle background. The
R12 value was calculated from a mean of four points arranged
around the cluster center at a distance of $2^{\circ}$. For all the
re-analyzed clusters the virial radius is less than $0.5^{\circ}$ so
this should eliminate any possibility of contamination from cluster
emission. ``Blank sky'' fields are then selected with N$_{\rm 21cm}$
and R12 values as close as possible to that of the cluster field to be
matched. The event lists are broken into 50 second time blocks and a
particle background calculated for each block using the flux measured
in the 12-15 keV band. A minimization is then performed by removing
(and adding) blocks from the pool until the mean values of the three
parameters is as close as possible to the values of the parameters for
the cluster field. A ``blank sky'' background is then constructed from
the selected event list blocks. This background needs to have an
exposure time significantly larger than that of the source to avoid
degrading the data quality with extra noise.

One other problem with these stacked backgrounds is that the ``blank''
fields inevitably contain large numbers of point sources.
Traditionally these are excised and large numbers of fields are
stacked in order to reduce the effect of the missing data. However it
should be noted that the effects of these excisions are clearly
visible in images of prepackaged backgrounds. Since the number of
``blank'' fields available is much reduced when trying to match a
background to a target observation, we instead mask out the regions
containing point sources from the source and background when creating
the spectra for fitting. In selecting ``blank'' fields to build a
background there is a strong constraint that they need to be as free of
point sources as possible so that as small an area of the cluster must
be masked from the spectral fitting. These constraints combined with
those of the background parameters (N$_{\rm 21cm}$ and R12) mean that
selecting observations for use in the matched backgrounds is a complex
process.

\begin{table*}
\begin{center}
  \caption{Properties of the observations used to construct the
    matched backgrounds for the four re-analyzed cluster fields.
    \label{tab2}}
\begin{tabular}{lccccccc}
\tableline\tableline
Cluster& Abell 1795& Abell S1101& Abell 1835 & MKW 3s\\
\tableline
OBSID:&0020540401& 0111550401& 0106660101& 0106660101\\
&0032140101& 0112630201& 0106660201& 0106660201\\
&0085170101& 0128531401& 0106660401& 0106660401\\
&0111550401& 0128531601& 0106660601& 0106660601\\
\tableline
N$_{21cm}$ ($10^{20}$ cm$^{-2}$)&1.16&1.82&2.32&2.32\\
\tableline
R12 ($10^{-6}$ s$^{-1}$ arcmin$^{-2}$)&1050.6&1051.1&1011.7&1011.7\\
\tableline
Exposure (ks)&101.5&145.0&192.6&124.5\\
\tableline
\end{tabular}
\end{center}
\end{table*}

\begin{figure*}
\plotone{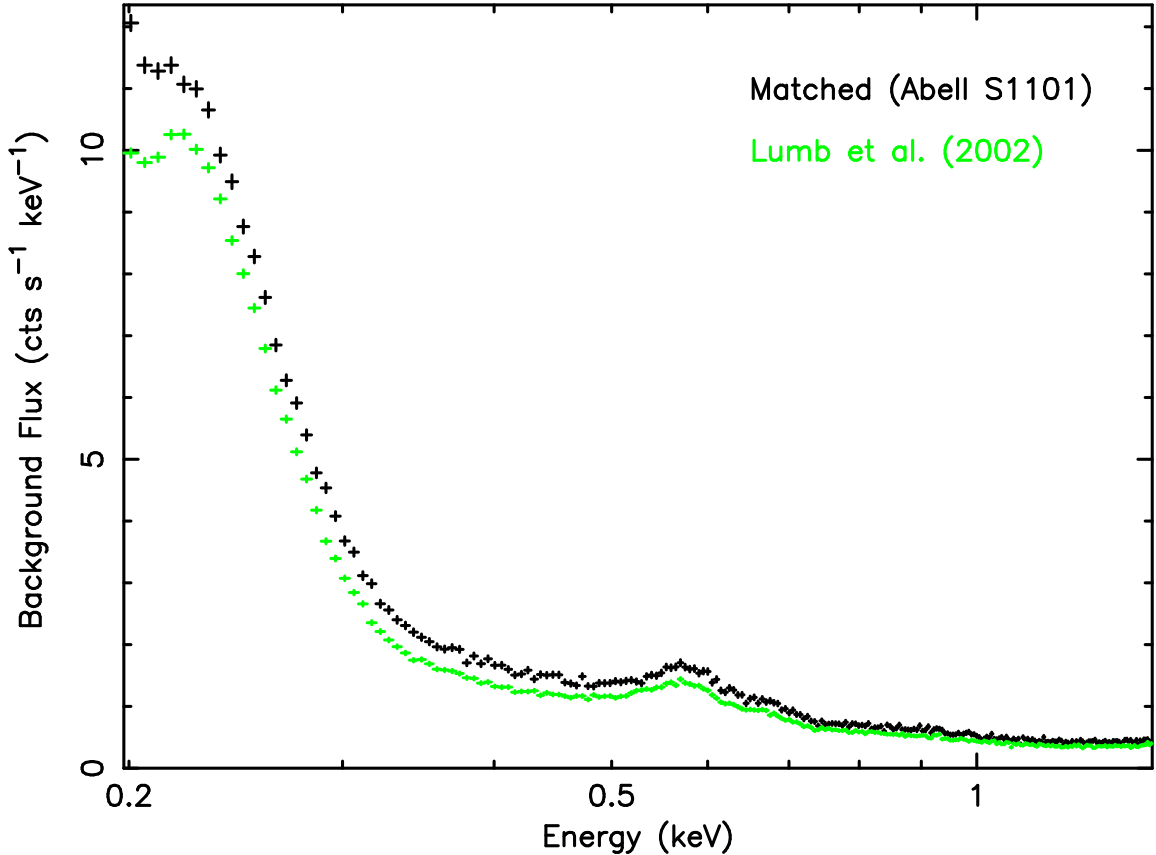}
\caption{Spectrum of matched background for Abell S1101 with the mean
  background of \citet{lumb02a} subtracted. The largest
  difference at lowest energy. Normalization is arbitrary.
  \label{fig5}}
\end{figure*}

We select four of the best candidates for a possible soft excess from
\citet{kaastra03a} to study the effects of using our matched
background technique. The systems are Abell S1101 (also know as Sersic
159-03), Abell 1835, Abell 1795 and MKW 3s. The properties for these
cluster fields are shown in Table \ref{tab1}. For each of these
clusters either the data used by \citet{kaastra03a}, or longer
exposures if available, were used. The data was reduced in the usual
manner and cleaned by performing a iterative 3-sigma clipping on the
12-15 keV light-curve to remove periods of high particle background.
Only EPIC-pn data was used since it is the instrument that receives
the highest count rate and if the effect (soft excess) is not
detectable in a single instrument then a detection using multiple
instruments would not be reliable given the uncertainties in
cross-calibration between instruments. Table \ref{tab2} lists the
observations used to construct the matched backgrounds and mean
properties of the matched backgrounds. Spectra were extracted in
annuli about the cluster centers with a minimum annulus size of 5
arcsec growing with radius to preserve the signal-to-noise. The latest
EPIC-pn response files, released in May 2005, with improvements to the
low energy response, were used for the analysis (see XMM-CCF-REL-189).
Background spectra were extracted from identical regions of the
matched ``blank sky'' background. The spectrum for each annulus was
fitted using a single temperature MEKAL plasma plus photoelectric
absorption model, using XSPEC, in order to obtain a measurement of the
hydrogen column.

\begin{figure*}
\plotone{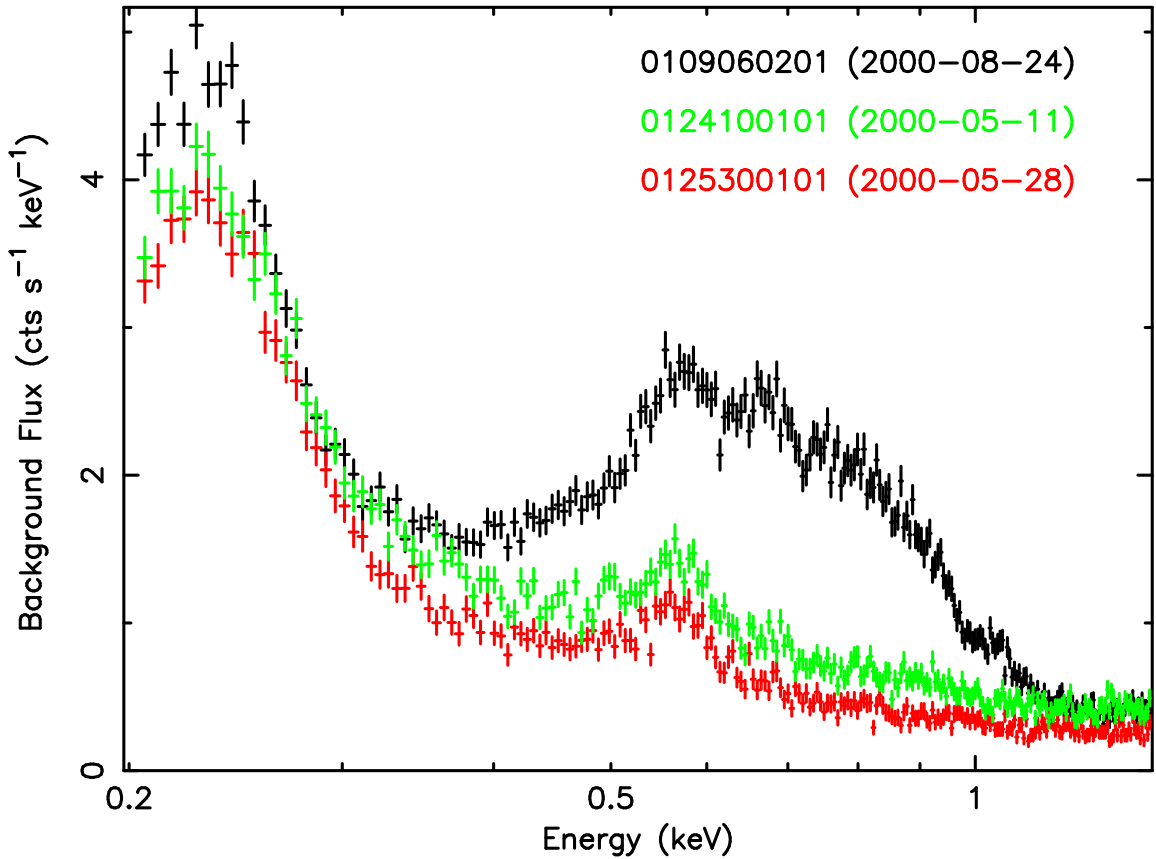}
\caption{Spectra of backgrounds taken from observations close in time
  to those of four of the clusters shown in Figure 6 of
  \citet{kaastra03a}: 0109060201 (MKW 3s and Abell 2052), 0124100101
  (Abell S1101) and 0125300101 (Coma). Observation 0109060201 (MKW 3s
  and Abell 2052) shows a large excess between 0.4 - 1.0 keV,
  particularly around the O VII and O VIII lines, presumably due to
  geocoronal or heliospheric charge exchange.  There is also some
  evidence of a slight enhancement for observation 0124100101 (Abell
  S1101), especially in the C VI lines. \label{fig6}}
\end{figure*}

The results of fitting the hydrogen columns for the four clusters
using the matched backgrounds are shown by the solid squares in Figure
\ref{fig4}. The errors are 1-sigma. The results from Figure 1 of
\citet{kaastra03a} are overlayed as open circles. The dashed lines
show the level of N$_{{\rm 21cm}}$ in each case. It can be seen that
the values measured using the matched backgrounds are roughly
consistent with the galactic N$_{{\rm 21cm}}$ value and are
significantly above the values measure by \citet{kaastra03a} using a
mean ``blank sky'' background.  \citet{kaastra03a} interpret the
significantly sub-galactic N$_H$ that they measure as due to excess
soft emission offsetting the absorption.  Our new analysis appears
to validate the hypothesis that the soft excess is due to incomplete
subtraction of the soft X-ray background since use of the matched
backgrounds cause the N$_H$ discrepancy to disappear.

To illustrate the difference between our matched backgrounds and the
mean background used by \citet{kaastra03a} we plot the spectrum of the
matched background for Abell S1101 along with the prepackaged
background of \citet{lumb02a} (Fig. \ref{fig5}). The backgrounds are
scaled so that they have the same flux in the 12-15 keV band to
account for differences in the particle background.  It can be seen
that the greatest difference in the backgrounds is at low energy, as
expected, with the difference falling rapidly with increasing energy.
This is due to the higher soft X-ray background in the fields used to
construct our matched background (since Abell S1101 has a high R12 and
low N$_{\rm 21cm}$) compared to the fields used to construct the mean
background of \citet{lumb02a}. If the mean background of
\citet{lumb02a} is used for background subtraction the soft background
emission seen in Figure \ref{fig5} will not be removed. This will
result in an apparent soft excess in the final spectrum.

\section{OVII emission}

\citet{kaastra03a} also report red-shifted O VII K$_{\alpha}$ lines
from a number of clusters in their sample which would also be an
indication for for the presence of cool gas. However for only two of
the clusters is the result really significant, Abell 2052 (at the 99\%
confidence level) and MKW 3s (at the 91\% confidence level), and this
significance hangs of the assumption that the emission is dominated by
the 574 eV resonance line and the 569 eV intercombination line. If the
emission is dominated by the 561 eV forbidden line though, the
significance is greatly reduced. \citet{kaastra03a} dismiss forbidden
line emission due to photoionisation but they do not consider emission
from heliospheric and geocoronal charge exchange. This is the result
of collisions between solar wind ions and neutral atoms from ISM
(heliospheric) and exosphere (geocoronal) and one of the strongest
expected lines is the O VII K$_{\alpha}$ forbidden line (see
\citet{snowden04a}, \citet{wargelin04a} and references therein). Solar
wind charge exchange emission is expected to vary significantly with
time and position on the sky. The geocoronal emission is generally
much weaker that the heliospheric emission except during periods of
enhanced solar activity.

At this point it should be noted that the observations of Abell 2052
and MKW 3s used by \citet{kaastra03a} were taken within a day of each
other (2000-08-21 and 2000-08-22) and are only slightly more than a
degree apart on the sky. To test whether these observations could
have been affected by significant charge exchange emission we selected
observations of non-extended sources taken shortly before or
afterward and pointed in a similar direction. Background spectra for
these observations are shown in Figure \ref{fig6}. It can be seen that
Observation 0109060201, which was taken shortly after the observations
of Abell 2052 and MKW 3s and is about 30 degrees on the sky away from
them, has a large excess of emission between 0.4 - 1.0 keV compared to
the backgrounds for the observations contemporaneous with those of
the clusters that do not show significant ``redshifted'' O VII
emission.  This correlation extends to the fact that the background
contemporaneous with the observation of Coma (the cluster for which
\citet{kaastra03a} detect the least excess O VII emission) also shows
the least evidence of charge exchange emission. This would seem to
strongly undermine the case for the emission being extragalactic in
origin.

\section{Discussion and Conclusions}

To investigate the cause of the soft excesses observed in several
clusters by \citet{kaastra03a} we have studied the correlation of
properties of the cluster fields with Kaastra's soft excess/deficit
measurements and find that the excess/deficit is correlated with both
galactic latitude and the soft X-ray (R12) background. From this we
infer that incorrect subtraction of the soft X-ray background is a
likely cause of the observed excess/deficits. To test this we
re-analyzed the data from four clusters for which \citet{kaastra03a}
measure significant soft excesses. Using backgrounds matched to the
properties (N$_{\rm 21cm}$, R12 and particle background) of the individual fields
we obtain fitted hydrogen columns consistent with the galactic 21cm
columns and considerably higher than those measured by \citet{kaastra03a}. 
We therefore conclude that the soft excess/deficit observed by 
\citet{kaastra03a} most likely the result of the field by field
variation of the soft X-ray background that remains in the data after
a mean ``blank sky'' background is subtracted.

\begin{figure}
\plotone{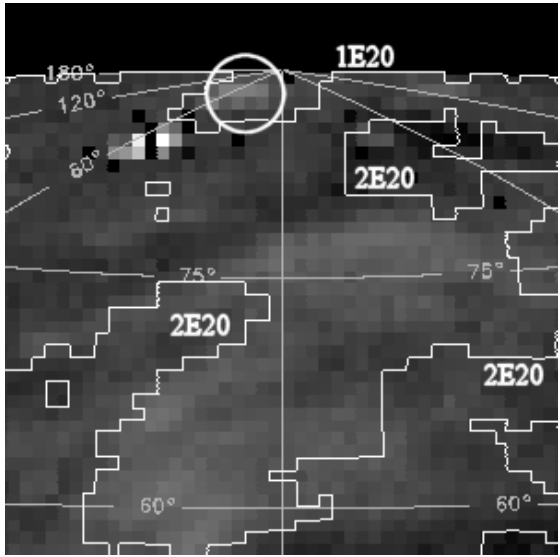}
\caption{The N$_{21cm}$ contours are superimposed upon the all-sky image of
the 1/4 keV X-ray in the region of the Coma cluster (at the center of the
$5^{\circ}$ diameter white circle).  The Coma cluster lies in a local HI
minimum, which is often correlated with a brightening in the 1/4 keV band.
The bright pixels at {\it l\/} = $55^{\circ}$ , {\it b\/} = $84^{\circ}$ is an
artifact.\label{fig3}}
\end{figure}

We have also examined the evidence for redshifted O VII lines
presented by \citet{kaastra03a}. This detection is significant only in
two clusters, Abell 2052 and MKW 3s, and only if the emission is
dominated by the resonance and intercombination lines. However we show
that the observations of these two clusters, which where taken within
a day of each other and are a degree apart on the sky, are likely
contaminated with heliospheric or geocoronal charge exchange emission
containing strong O VII forbidden line emission. Given the correlation
between our expectation of charge exchange contamination and the the
detection of ``redshifted'' O VII emission by \citet{kaastra03a} we
conclude that the case for the emission being associated with clusters
is very weak.

The one source that several authors agree upon as having an apparent
X-ray excess is the Coma cluster, so we examine whether this is truly
evidence for a {\it WHIM\/} component in cosmic filaments
\citep{bonamente03a, finoguenov03a}.  Even on the R12 map, one can see
an enhancement in this region of the sky, suggestive that it is due to
the Coma cluster (Fig. \ref{fig3}).  However, there is a local minimum
in the 21 cm column density at Coma \citep{dickey90a, hartmann97a},
and this was certainly not caused by the Coma cluster.  A local
minimum in the HI sky would permit us to see the soft X-ray emission
from the Galactic halo more readily, causing a brightening in the soft
X-ray map of the sky at that location.  It is extremely difficult to
separate a Galactic (halo) brightening in R12 from that associated
with the Coma cluster.  There are clear examples of brightening in the
the R12 flux at low 21 cm column regions, such as in the Lockman hole
\citep{snowden94a}.  Furthermore, the 21 cm column density is so low
in this direction ($<$ 10$^{{\rm 20}}$ cm$^{{\rm -}{2}}$) that it is
poorly known as the various instrumental corrections (from sidelobes,
etc.)  become a significant fraction of the signal
\citep{hartmann96a}; the true 21 cm column may be lower than the usual
values quoted.  Further complicating the analysis of Coma is that its
extremely low N$_H$ would make finding ``blank sky'' fields with
similar properties very difficult. Therefore, we are cautious about
claims that the Coma cluster possesses a soft X-ray excess and we note
that \citet{arabadjis00a} were able to fit a free-free emission model
to Coma ({\it ROSAT PSPC\/} data) without an additional soft excess,
but they required a Galactic column density (6$\times$10$^{{\rm 19}}$
cm$^{{\rm -}{2}}$) lower than the values of \citet{hartmann97a}
(9$\times$10$^{{\rm 19}}$ cm$^{{\rm -}{2}}$).

Despite our concerns on the reality of a soft X-ray excess from
clusters of galaxies, it might be possible to isolate this component
spectrally, which should be possible because this soft excess is
rather bright.  If one could show redshifted OVII line emission from
the outskirts of a cluster that was not blended with the Galactic
feature (or contaminated with charge exchange emission), it would
constitute strong evidence for the {\it WHIM\/} around clusters.
Currently, the OVII line is blended with the Galactic OVII line for
low redshift clusters and does not appear to be present in the
moderate redshift cluster (Abell 1835) whose spectrum would imply a
soft excess (f$_{{\rm N}}$ $<$ -0.3).  The use of smaller, higher
redshift clusters would also help to isolate the soft excess because
one could take a local background from the same field of view as that
used to image the cluster.

\acknowledgments

We would like to thank Jimmy Irwin, Renato Dupke, Eric Miller, John
Arabadjis, Wilt Sanders, and Steve Snowden for comments and advice.
We would also like to thank the referee for his useful comments. We
acknowledge support from NASA grants NAG5-10765, NAG5-13137, and
GO1-2147X.  This research has made use of data obtained from the High
Energy Astrophysics Science Archive Research Center (HEASARC),
provided by NASA's Goddard Space Flight Center. We acknowledge the use
of NASA's {\it SkyView} facility (http://skyview.gsfc.nasa.gov)
located at NASA Goddard Space Flight Center.


\end{document}